*Philosophie ist die Wissenschaft aller möglichen*
*Dinge, wie und warum sie möglich sind.*
Christian Wolff[1]

*Chapter 9*

# THERE MUST BE ENCAPSULATED NONCONCEPTUAL CONTENT IN VISION

## *Vincent C. Müller*

Department of Social Sciences, American College of Thessaloniki

### INTRODUCTION

In this paper I want to propose an argument to support Jerry Fodor's thesis (Fodor 1983) that input systems are modular and thus informationally encapsulated. The argument starts with the suggestion that there is a "grounding problem" in perception, i. e. that there is a problem in explaining how perception that can yield a visual experience is possible, how sensation can become meaningful perception *of* something for the subject. Given that visual experience *is* actually possible, this invites a transcendental argument that explains the conditions of its possibility. I propose that one of these conditions is the existence of a visual module in Fodor's sense that allows the step from sensation to object-identifying perception, thus enabling visual experience. It seems to follow that there is informationally encapsulated nonconceptual content in visual perception.

### 1. MODULES IN FODOR

Fodor has proposed that we look at the mind as built up partly of modules that serve particular processing tasks: "modular cognitive systems are domain specific, innately specified, hardwired, autonomous, and not assembled" (1983, 37). Fodor distinguishes "transducers, input systems, and central processors, within the flow of input information" (1983, 41f). Transducers provide a "distribution of stimulations at the 'surfaces' (as it were) of the organism", so they are merely translating the external stimulations into output that can be processed by the organism. The "input analyzers" process the output of transducers and deliver representations that are "an arrangement of *things in the world*" (1983, 42). It is these analysers that are thought to be modules and consequently the output of these modules also marks the distinction between perception and cognition in central processes.

---

[1] "Philosophy is the science of all possible things, how and why they are possible." Wolff 1713, Vorbericht, §1.

The main characteristics of input modules are absence of voluntary control by the subject and the informational encapsulation of the modules from top-down cognitive processes. Fodor starts his investigation by saying that modules are "basically, … a reflex" and, characteristically, our reflexes are not subject to volition: You cannot suppress the blinking reflex even if you have very good reasons to do so (1983, Dedication & 71). The same applies to the output of perceptual modules, "You can't help hearing an utterance of a sentence (in a language you know) as an utterance of a sentence" (1983, 53) because the relevant input module has already analysed it as such, whether you want it or not. Also, you cannot help seeing certain perceptual illusions "wrongly", even if you have the necessary knowledge: they persist despite knowledge of the individual that they *are* illusions (Müller-Lyer, phoneme restoration, the apparent movement induced by pressing your eyeball, etc. [1983, 66-70]). This seems to show that what is happening in these illusions is mandatory processing within the module.

Crucially, the modules have access only to information from sensation and inside the modules themselves, they do not have "bottom-up" access to our other mental states, concepts, beliefs or theories, so there is no top-down influence. Also, the content of the module is not directly accessible to central processes, only their output is. In that sense the modules are informationally encapsulated. "The informational encapsulation of the input systems is, …, the essence of their modularity." (1983, 71; cf. 2000, 63) Modules have "shallow" outputs: "In general, the more constrained the information that the outputs of perceptual systems are assumed to encode the shallower their outputs, the more plausible it is that the computations that effect the encoding are encapsulated." (87) A module is computationally autonomous, i.e. does not share resources such as memory, attention or judgement (Fodor 1983, 21). Finally, input modules are domain specific (highly specialised on particular stimuli) and remarkably fast, while central "non-automatic" processes under voluntary control are slow.

A feature that is crucial for the argument proposed below is that modules are innate; not in the sense that a particular faculty is innate to a particular person, but in the sense of an instinct, being equally innate to all members of a species (Fodor 1983, 20). Given the types of arguments used by Fodor (and here below), it is to be expected that perceptual modules are present in all animals that can see as we can, and incarnated in similar ways in otherwise similar animals, e.g. in all mammals. Unlike in Chomsky's account (Chomsky 1980, 3ff) and the classical rationalist tradition, one should not think of the content of the innate modules as a body of propositional content or as a theory but rather as a structure, a faculty. Fodor is thinking "not of an innately cognized rule but rather of a psychological mechanism – a piece of hardware, one might say –" (1983, 8).

There is a fair amount of debate as to how much of cognition is modular. Fodor 1983 had distinguished between modular input systems and central processes, the later Fodor (2000) argued that the limits of computational cognition are more narrow than some of his followers had thought and, given the globality of most cognition, the thesis that most or all of cognition is modular must be false (2000, 55ff). The question thus remains how much of the modularity thesis can be salvaged for input systems and how much processing takes place inside the module. Fodor was initially quite willing to include a fair amount of processing into the perceptual module. He discusses the conceptual hierarchies in the theories of E. Rosch etc.

who point out that in a series of terms in a hierarchy that all apply to the same object, (say, *poodle, dog, mammal, animal, physical object, thing*) we are more likely to attribute terms from one salient level - in this case we would typically call something a *dog*, this is thus the "basic category". Fodor explains this fact by suggesting that this is the most abstract level where the objects that fall under the category still look similar - dogs look similar, mammals do not. He concludes that the input module performs object identifications and "So, the suggestion is that the visual-input system delivers basic categories" (1983, 97 & fn. 34). Putnam interpreted this as saying that the module has statements, such as "That is a dog" as its output (Putnam 1984, 411). Even if a *statement* is not the output, the problem remains how the module can have the concept "dog" for basic categorisation. It cannot be innate, it seems, and it is not clear that it could be constructed inside the encapsulated module without top-down influence. It is also obscure why this concept should be constructed and not another, like *poodle* (under suitable environmental conditions). In fact, it is not clear why certain other concepts like "my king is in check" should not also be part of the visual module. At the same time, Fodor talks as if modules should have *only* content that is innate, not changed or acquired by learning. As we mentioned, he stresses that the Müller-Lyer illusion shows that learning does *not* help - the illusion remains despite my knowledge that it is an illusion. It seems fair to conclude with Putnam "that no presently surveyable limit to the class of 'observation concepts' is set by the structure of the visual module" (1984, 414). In other words, there is still a question as to at what level exactly is the output of the visual module, if it exists.

Given this situation, I propose to step back and see whether there are other arguments that support the need for modules and how much content in the modules these arguments would require.

Fodor himself says "I have not yet given any arguments (except some impressionistic ones)" (1983, 73) *for* the thesis that perceptual systems are actually modular. This is after he has mentioned: a) the persistence of many illusions despite our knowledge that they are illusions, and b) there has to be more in the input systems than what the organism already expects (1983, 68-70). In the process, Fodor only offers defensive arguments to the effect that traditional arguments *against* modularity (against cognitive encapsulation) are less convincing than commonly thought (65f.). He concedes that these arguments do show a top-down causal relevance of knowledge to perception, but denies that they imply top-down relevance to perceptual modules (74). Given this discussion and the scarcity of empirical support for the modularity thesis, we are still in need for a positive argument for the existence of such modules.

## 2. Sensation, Perception and Experience

Before I proceed to the main body of the paper, I would like to propose some terminological distinctions: Let us divide the cognitive process that results in vision into three parts: sensation, perception and experience.

Let us call visual *sensation* all input processes that lead to the formation of the retinal image. This begins with the mechanical processing of light that reaches the cones on the retina and is transformed into internal signals for further analysis in the neurons. I would include processes that compute differences, changes and patterns in light intensity by locating and coding individual intensity changes. (Marr's *raw primal sketch* that provides information about edges, bars, blobs, boundaries, edge segments etc., is an example of sensation.) Sensation is not accessible to introspection. Sensation could be attributed to a blind person who's eye and visual nerve is intact, but who suffered severe brain damage and shows no behavioural response to visual stimulation.

The processes that transform sensation to a representation that can be processed by cognition are called *perception*. (One example are Marr's various grouping procedures applied to the edge fragments formed in the raw primal sketch. They yield the *full primal sketch*, in which larger structures with boundaries and regions are recovered. Finally, in Marr's model of vision the product of perception is the *21/2D sketch*.) Whether or not perception involves concepts or is informationally encapsulated is an empirical matter.[2] Perception is what the blindsighted person has, who can identify objects and act on his perception but is not aware that he does. Similarly it is what a normal human does in certain high-speed situations (and in peripheral vision) without awareness of what he/she is doing. Perception is not in principle inaccessible to introspection.

All subsequent visual processes fall under *visual experience*. Experience is conscious and shows no informational encapsulation, so it is accessible to introspection and influenced by our beliefs and desires and it directly influences our beliefs and desires. It involves a strong conceptual and knowledge component (thus allowing us to see the red blob in the visual field as Bob climbing the dangerous mountain). Experience is significantly "poorer" than perception, in the sense that the interpretative processing leaves out many features of the perceptual input. Experience has a phenomenal quality, that is, there is something "what it is like" to have an experience.

What is contentious about this terminological proposal is whether there is a sensible distinction between sensation and perception. Given the criteria of consciousness and belief generation, there seem to be fair grounds to distinguish visual experience from perception and sensation. In a similar vein, Tye (2002, 447) distinguishes between the phenomenal content of a scene, which includes colours, shapes, and spatial relations obtaining among blobs of parts, and the semantic or conceptual content of a visual scene (the fact that it contains, say, a tiger). Fred Dretske (1993) distinguishes "thing-awareness" from "fact-awareness" and also (Dretske, 1995) a "phenomenal sense of see" from a "doxastic sense of see". The first parts of the aforementioned pairs correspond to perception, the second parts to experience. Given that the content in experience is the content of judgements and beliefs, it is clearly a conceptual content, while perception and sensation do not a priori require conceptual content. If the variation of Fodor's modules proposed below is correct, informational encapsulation could provide the criterion for making the distinction between perception and experience - along the lines of Raftopoulos/Müller 2003b who propose to identify non-conceptual content with informationally encapsulated content.

---

[2] The distinction drawn here is probably not identical with that between low-level and mid-level vision. Low-level vision involves information such as surface shading, texture, edges, color, binocular stereopsis, size, and analysis of movement. Mid-level involves spatiotemporal and size information, color and its properties, as well as some depth-encoded surface representation of the layout (see Marr, 1982; Pylyshyn, 1999; Raftopoulos 2001, for discussion and further references). None of the two involves identification of objects.

# 3. How Is Vision Possible?

## 3.1. The Grounding Problem

I would like to suggest that there is a "grounding problem" in vision and that, given this problem, the best explanation for the possibility of vision involves the postulation of visual input modules. The so-called "grounding problem" is originally a problem in computing, specifically in artificial intelligence.[3] For example, how could you make a computing machine that has vision? Imagine the first step: light that passes some camera lenses and meets an array of sensors. Each of the sensors will produce some output in a form that can be computed by the remaining mechanism (this is true even if you are allergic to Cartesian dualist legacies) - in a conventional computing machine this is a sequence of on/off electrical charges (bits). This sequence contains the symbols that the system can manipulate. Now, how does this sequence of bits, this output of sensation, ever become more then just a sequence but a sensation or representation *of an object*? How can these symbols be grounded in the world and become symbols of something? Seen from outside the system, the symbolic output of sensation can be interpreted as a meaningful symbol of what was perceived, but how does the output become a symbol that is *meaningful to the system*? If the system manipulates symbols, the meaning cannot come from further symbols. As Harnad asks: "How can the meanings of the meaningless symbol tokens, manipulated solely on the basis of their (arbitrary) shapes, be grounded in anything but other meaningless symbols?" (Harnad 1990).

It is useful to explain the problem in a slightly different form, still referring to symbols in computers. In his celebrated "Chinese Room Argument", John Searle (1980) tries to show that computers cannot think because they cannot understand anything. He imagines himself doing the work of the central processing unit (CPU) in a computer that analyses the content of English texts:[4] He sits in a room where he receives input in a symbol system that he does not understand (Chinese) and follows instructions in a manual written in a symbol system that he does understand (English), on how to manipulate the Chinese symbols and which to produce as output. Searle concludes that he, the man in the room, would never learn Chinese by doing these symbol manipulations, that he does not and will never understand the symbols that are the input and output of his room. This much has been granted by all critics, as far as I can see. Searle concludes by analogy that a computer is just like the a man in the Chinese room, receiving input that he does not understand, manipulating it and producing output that it does not understand – so it will not understand anything ever. This analogy is where the critics have taken him to task in various ways (though unsuccessfully, as Searle 1999 still believes).

Amongst the objections that are already discussed in the original argument is the "robot response" to the effect that the computer should be equipped with "sense organs", such as cameras, microphones, etc. This equipment, so the response goes, would provide causal interaction with the environment that would allow the symbols in the system to acquire meaning. Searle responds to this by saying that whatever input into the Chinese room, be it from cameras or anything else, is "just more Chinese" to the man in the room. It would not help him to understand the symbols, he would not even understand that this input is from the

---

[3] Its first extended discussion and proposal for a solution (via complex isomorphisms) is Hofstadter 1979.
[4] So his argument initially just applies to Van-Neumann-machines, but he thinks that it can be expanded to neural networks, too (in the "Chinese gym").

"sense organs" (or from an outside world, for that matter). This response seems correct to me.[5]

The structure of our grounding problem can now be put in terms of the Chinese Room: What would the man in the room need in order to learn about the outside world and thus acquire meaningful symbols? If there is a way out of the Chinese Room, a way for that system to relate to the outside world and thus give its symbols meaning, it must involve the whole computing system with its causal relation to the outside world (person in the Chinese room, instruction manual, sensory machinery - processor, software, further hardware - roughly: central systems, perceptual analysis, sensual apparatus).

So, what would a human being need?[6] What allows the central system to learn from sensation? We think of the problem as "psychological systems whose operations 'present the world to thought'" (Fodor 1983, 101), and ask for an explanation of how this is possible. Before we proceed towards solutions, let me stress that this grounding problem for vision in humans does not occur each and every time a perception takes place; rather it is an ontogenetic problem, the question how a particular system can develop the ability to see: How is it possible that a human infant opens his/her eyes and sees enough to learn? Again, given that we know that children learn to see, there must be a solution to this grounding problem for humans.

## 3.2. Epistemological Bootstrapping: Locke vs. Leibniz

To get a better idea of the shape the solutions to this kind of problems should take, let us take a brief look at an analogous historical debate that has produced significant insights some 300 years ago. Our problem is one of a number of problems of *beginning*, from the classical epistemological problem of how we can know anything at all (probably still best analysed by Hegel) to the "bootstrapping" problem of a computer: you have to build-in some routines such that it can start up by finding the mass-storage (e.g. hard-disk), read its contents and thus load the "operating system" - in a conventional computer this is done by the built-in BIOS (basic input-output system).

Concerning the question of how ideas enter the mind, on the one hand there was classical empiricism, characterised by Locke's remarks like "The Senses at first let in particular *Ideas*, and furnish the yet empty Cabinet" (Locke 1689, I, 1, §15) or "Let us then suppose the Mind to be, as we say, white Paper, void of all Characters, without any *Ideas*;" (II, 1, §2) On the other hand, the rationalists asked whether it is possible that the mind presents itself as a complete blank to the world. G. W. Leibniz wrote a detailed response to Locke in his *Nouveaux Essais* and points out: "… cet axiome reçu parmi les philosophes, *que rien n'est dans l'âme qui ne vienne des sens*. Mais il faut excepter l'âme même et ses affections. *Nihil est in intellectu, quod non fuerit in sensu,* excipe: *nisi ipse intellectus.*" Accordingly, he thinks the mind is not a blank but itself already contains some ideas even though we are not aware of it: " … les idées sont en nous avant qu'on s'en aperçoive" (Leibniz 1705, I, 1, §8)[7] and so he goes on to argue for the innateness of certain ideas (of necessary a priori truths, in particular).[8] It is these innate ideas that allow us to fill Locke's "cabinet" with ideas. Leibniz points out that for what is inside the soul we cannot use introspection, since there is no reason to assume that we are can be aware of it; instead, we need to look at what is possible, and why (as our motto by Wolff suggests).

---

[5] In fact, the situation is worse for computers: Unlike Searle, who understands the English of the instruction manual and might understand what he is doing, the computer does not even understand the software, it just acts mechanically on the binary input, carrying out an algorithm consisting of elementary operations on the memory registers (read, delete, write, compare, copy, …).

[6] Given that Searle thinks that humans are machines that can think, he also assumes that there is a solution to this particular kind of grounding problem for humans. The magic connection he eventually proposes is "intention", which non-biological machines supposedly cannot possess. I will indicate below that this is just one of several conditions for the possibility of perception.

We need to keep in mind that what was at stake between empiricists and rationalists was the foundation of knowledge, not the investigation of cognitive processes. Accordingly, Leibniz and Kant do not discuss what would be necessary to have any experience at all but what is needed to make knowledge possible. Having said that, the structure of our argument will have to be the same as that of all rationalists and of Kant. Kant (1781) stressed that certain ideas, particularly space and time are needed for the acquisition of empirical knowledge, but cannot be acquired through experience, so the question arises how knowledge is possible. The answer is that these ideas must exist prior to experience, built into pure reason.

One way to make this rationalist point would be to say that Hume has not been sceptical enough about what we could learn, which concepts we could form, just from experience. If it was just for impressions on a tabula rasa, we would never learn anything at all - just like the computer would never acquire meaningful symbols. Whether you want to explain how humans are so good at learning a specific thing (Chomsky) or how they start to see anything at all, in either case you have to move away from pure empiricism. So, in order for perception to be possible, what is it that needs to be innate?

## 4. THERE MUST BE MODULES

### 4.1. Modules

So, what are the conditions under which the grounding problem can be solved? How is it possible that sensation can become perception and experience? Since we are to understand this as a problem of ontogeny, what we are asking is what has to be innate for the human to arrive at perception and experience. Keeping in mind the Chinese Room analogy, I would suggest that we need three things to be innate: A) the will to learn and refer to an outside world, B) awareness which input is from which sense organ, C) structures of the input from sense organs that allow the system to identify and re-identify objects. I do not argue for A) and B) here – though it will become apparent that B) is required for experience.[9] I maintain, however, that C) is necessary and that, therefore, Fodor's perceptual modules are necessarily innate.

---

[7] In English: "… that received axiom of philosophers that *nothing is in the soul that does not come from the senses*. But we must except the soul itself and its affectations. *Nothing is the intellect that was not in the senses, except: intellect itself*", "… the ideas are inside us before we observe it"

[8] On some views (quoted in Fodor 1983, 11) Locke did not mean to exclude the existence of natural faculties and innate mental powers by his remarks, assuming that the mind has the necessary apparatus to acquire ideas through experience. I tend to think this is untenable in the face of Locke's statements about white paper and empty cabinets that seem to deny precisely Leibniz' point that that mind is "already there", having a structure.

[9] The importance of will for a Fodorian symbol system is discussed in a separate paper on what I call the "information paradox".

Are there more necessary features? Apart from obvious additions such as the actual existence of sense organs, it appears essential to human learning that we have several sense organs that produce input from the same object. I suspect that a system with only vision (seeing the world like in Plato's cave) would be severely limited in its abilities to form representations of the world. In any case, it does no harm to the present argument if there are *more* necessary conditions for perception and experience; what we need to maintain is that at least C) should feature in the list.

So, what is required is enough structure from the sensory output to build elementary representations that allow for learning, meaning that the subject must represent objects as distinct from one another. All processes that are necessary for this step are part of the module.[10] This does not imply that these objects must be represented as having particular properties, but just to be identified, distinguished from one another (so that, for example, one can see that one thing is moving with respect to another or with respect to the subject). So, in order for the module output to be not just "noise" or "more Chinese", it must already exhibit structures that allow further processing. *Just* structure is not sufficient, however. Consider the analogous problem of astrophysicists that try to find out from the analysis of radio signals from outer space whether there is extraterrestrial intelligence. If the signal exhibits some structure, some pattern, then the question arises whether this was caused by some intelligence. Even if a pattern may, at some point, be sufficient to be considered evidence that there is indeed some intelligent being out there, there is no way that the structure could be decoded to mean something like "Hello there, we are the Argonauts on planet Iolkos – come to visit us at co-ordinates 08/15." (Imagine the astrophysicists would pick up a signal from something like that of a human TV station. Consider the analogous problem of putting a meaningful message on human-made satellites that leave our solar system …) In order to decode a message we need to know *something* about its origins and be causally related to those. In the case of vision, I propose that what is required is a representation that is *already* spatial, where areas and edges could appear as such (not encoded) and thus allow simple object identification. So we need to imagine this stage not like the analysis of radio signals form outer space or our local TV station but more like looking at a very badly tuned TV screen, black & white, with lots of noise, where you can just about make out something - but which is more than just encoded signals.

I have argued in earlier papers with Th. Raftopoulos that there is actually philosophical and empirical evidence that such a mechanism exists. We suggest there that there are causal mechanisms that allow content to be retrievable from the perceptual scene in a bottom up manner that involves nonconceptual content. The empirical evidence suggests that there are mechanisms of eye-fixation for tracking objects that work independently of any knowledge about features of these objects. We suggest that "the individuation is accomplished by opening an object-file fixing the object to which the demonstrative refers and allowing its tracking. … All the elements of the object-file that individuate an object can be retrieved directly from the scene in a bottom-up manner by means of the mechanisms of early vision (spatial position, colour, local content)" (Raftopoulos/Müller 2003a).

---

[10] In this discussion, I ignore issues of attention, talking as if what meets the eye also enters processing. I hope this is not a problem and that the (partly conceptual) procedure of attention can be ignored – in fact I suspect that there must be a similar grounding problem and nonconceptual beginning there, too. In any case, as mentioned under A) above, a "will" to refer to the outside world is a necessary part of the story.

What is required to achieve this representation goes well beyond early sensational processes like retinal filtering and sharpening, it demands a representation that is already spatial, where areas and edges appear as such; so we require the level of Marr's primal sketch, providing zero-crossings, discontinuities, edge segments, boundaries and groupings. But even this does not give us objects yet, we need rules like the ones Raftopoulos (2001, 429) mentions for the production of the 2 1/2 D sketch from the 2 D stimulation: "local proximity" (adjacent elements are combined), "closure" (combination of two edge-segments), "continuity" (smooth continuity of objects is assumed), "compatibility" (similar elements at similar spots are matched) and "figural continuity" (the figural relationships used to eliminate 'wrong' matches). These would be sufficient for the identification of objects. These principles should not, in my view be regarded as statements or knowledge about the structure of the world. Instead, they are tools, mechanisms necessary to construct any access to the outside world in the first place (which is why it constitutes an evolutionary advantage to have these). It is at this stage that objects can be distinguished for further processing, so I would expect the content of the visual perceptual module to reach somewhere in the region of the 2 1/2 D sketch. If this module were not innate, we would never learn to see.

Accordingly, the visual phenomena that indicate what is happening in the module are the gestalt phenomena (and the associated illusions) such as Marr's triangle, the Necker cube, pop-out phenomena, Wertheimer groupings (where in a matrix of dots, one sees the dots forming horizontal or vertical lines, depending on distance between the dots). This point might also be illustrated by a modified version of the Müller-Lyer illusion:

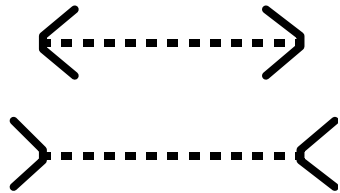

*Fig. 1:* Müller-Müller-Lyer Illusion

In this version the lines are dotted. Not only do the two lines appear of different length (as in the classical version) but the dots also appear to have different spacing and we cannot help seeing a line where there are just dots. These illusions occur despite a) an equal number of dots, b) equally spaced dots and c) "blunt" arrows in both cases providing a short vertical line at exactly the same width.

It is no accident that the evidence Fodor presents typically involves visual illusions that illustrate such processing - and the absence of voluntary control over these processes. I would agree that voluntary control would show that a process is not part of the module (for lack of encapsulation); the absence of voluntary control however, is not sufficient to show a process to be part of the module. Consider for example, the gestalt switches in the in duck-rabbit images or the cases where one can, all of a sudden, see a face hidden in an image but can then not voluntarily cease to see it - these phenomena clearly involve learned concepts, so they cannot be part of the innate module at issue here.

## 4.2. Nonconceptual Content

This level of processing is what is necessarily built-in into the visual module. The module is thus *informationally encapsulated*, at least initially. In other words, the information outside is not available to the module or, what is the same thing, that the information flow is only in one direction from the module to processes outside the module; there is no top-down flow. Whether or not the content is accessible to introspection is irrelevant for the present purposes - but all the evidence suggests that it is not.

The module presents or represents the world as being in a certain way and thus has normative conditions of correctness, though not the standard structure of discursive judgements. Is it conceptual? A conceptual content should *lead to* a belief but, it is characteristic of the processing stages under discussion here that they do not. As Gunther (2003, 10) pointed out, it is characteristic of illusions like Müller-Lyer that perception does not lead to belief (the false belief that the lines are of different length) if the person has the appropriate background, so if the workings of the illusion are part of the perceptual module, then this is *nonconceptual.*

Also, whatever its specific processing mechanism might be, given that this module is innate, not learned, it cannot be conceptual in the sense that it would depend on a person's acquired concepts, be they connected to the meanings in a public language or not. Quite the inverse, in fact: the possibility for the person to acquire concepts depends on the visual module (and the other perceptual modules). If the module remains encapsulated throughout a person's life, then a person's acquired concepts cannot enter the module at any stage and neither can explicitly formulated conscious beliefs and desires. (Of course, I am not claiming that the module is nonconceptual in the sense that it cannot be described conceptually - after all, this is what I am trying to do presently.) So, this supports the proposal in Raftopoulos/Müller 2003b, that nonconceptual content is "the output of bottom–up, nonsemantic processes of perception that are cognitively impenetrable". Given that our module is postulated a priori and thus remains minimal, it may well turn out that further nonconceptual processes are happening "after" the module.

Even if we grant that the content of the visual perceptual module is initially nonconceptual, it is conceivable that it can be influenced by conceptual knowledge once the person has acquired such knowledge (as in some computers where one can change BIOS settings once they are up and running). How much this is the case is an empirical matter. The evidence from visual illusions suggests that humans are fairly similar in this respect and that many aspects, like the ones Fodor stressed, cannot be influenced by beliefs and desires. Our visual experience differs only in those areas where conceptual knowledge clearly plays a role (like the duck-rabbit, the face recognition problems etc.). I would suggest to assume that those parts of visual perception that, according to empirical research, are not changed by higher cognitive processes are part of the first visual module. Also, despite ontogenetic stability, the content of the module might change in the course of evolution of the kind (phylogenesis) - this remark is only as paradoxical as any claim that the offspring of one species always belongs to that species but evolution is still taking place (the "paradox of evolution").

The argument that has emerged is an a priori argument, in particular what is now commonly called a "transcendental argument" in the tradition of Kant. In that tradition, our investigation proceeded a priori in the sense that we need only the actual existence of vision as an empirical datum. It lies in the nature of this method that we can hope at best to find characteristics that are necessary for vision, so we might miss out on features that are present even though they are not necessary. Accordingly, what we specify is a *minimal* account of what is innate. It may turn out that more than what is required on a priori grounds is *actually* innate – rather than learned from the innate structures plus environmental stimuli. Also, the modular account will say very little about the neuronal realisation of the module, given that a computationally identical mechanism could be physically instantiated, (hard wired in the neurones) in many different ways. In other words, even if we could identify on a priori grounds which output is that of an innate system, this would not tell us anything about the internal workings of that system. In particular, even if we assume it to be computational, that is algorithmic symbol-manipulation, we do not know its organisation and the simplicity of its parts. We might be able to tell from the relation between input and output *what* it is doing, but not *how* it is doing it. This remains a question for empirical research in the neurosciences.

## 4.3. Second Thoughts

At this point, a general concern about modularity might be raised. Whichever processing we postulate to take place in the module could in principle be done by central (general purpose) processes also. This should be clear from our radio-signals analogy: if the module can do a useful decoding that allows understanding, then there is an algorithm that does the job, so why should the astrophysicists not be able to find it? Fodor provides a hint in the right direction here: "In particular, the constructibility *in logical principle* of arbitrarily complicated processes from elementary ones doesn't begin to imply that such processes are constructible *in ontogeny* by the operation of a learning mechanism of a kind that associationists would be prepared to live with. This is a point about which I suspect that many contemporary psychologists are profoundly confused." (Fodor 1983, 34). We need to assume a model that allows the organism ontogenetically to form the right (computational) mechanisms. What is necessary *in the mind* to do this? Is there any reason to assume more than just central processes?

Now, what would I need in order to decode the signals from a TV station by hand? (Assuming I do not have a TV that does the job for me.) I suggest that I would need to A)' want to do find out what these signals encode, B)' know that this is a sequence of images and C)' know some of the properties of the images. Part C)' is precisely what the module does, as suggested above. In other words, the module has some built in structures that reflect knowledge about the objects from which the sensations causally originate (edges, continuity, …). If no such structures occur, the input will be "just noise, or "just more Chinese" to the central system, event though it could, in principle decode the signal. Saying that it could in principle decode the signal is like saying that I could, in principle, understand any language: Yes, provided that I know something about the "speakers" of the language and the world they live in and stand in causal relation to the latter. Now, given the visual module, we do have a mechanism that has the first meaningful (referring) symbols, a mechanism that does not itself rely on further symbols.[11]

---

[11] In the case of computers (algorithmic symbol manipulating machines), we clearly talk about symbol tokens, in the case of humans it may be better to speak of "representations" instead, but the problem remains the same: how can a representation become a representation *of something* for the subject?

One related concern: Fodor ridicules an evolutionary "poverty of the stimulus" a priori argument for full modularity by Cosmides and Tooby to the effect that "[It] is in principle impossible for a human psychology that contained nothing but domain-general mechanisms to have evolved …" (1994, 90; quoted in Fodor 2000, 65). Fodor responds that "poverty of the stimulus arguments militate for *innateness*, not for *modularity*. … You can thus have perfectly general learning mechanisms that are born knowing a lot, and you can have fully encapsulated mechanisms (e.g. reflexes) that are literally present at birth, but that don't know anything, except what proximal stimulus to respond to and what proximal response to make to it." (2000, 68f.). First, as his response makes clear, even the innate reflex has a structure and "knows" quite a lot - though clearly not as much as Fodor would like it to know. Second, remember that Searle in the Chinese Room knows quite a lot about the world and the origins of these symbols, but still does not manage to decode anything. The argument here is not a poverty of the stimulus argument for innate content, since it does not argue through the poverty of the stimulus (the stimulus is sufficient if properly analysed) but for the inability to *be* a stimulus. As we said, the output of the visual module must already have spatial features, originating from features of the world, be a spatial stimulus, otherwise it would be "just more Chinese".

## 5. CONCLUSIONS

I have tried to indicate why there is a grounding problem for perception and that we know there must be a solution. Furthermore, I argued that Fodor's innate perceptual modules do provide part of what is necessary for such a solution, so we should postulate their existence. If this move is successful, we now have an a priori argument for perceptual modules, at least in a minimal form. Finally, I indicated how such modules must contain nonconceptual content. I thus provide some material for the explanation of the steps from sensation to perception and, ultimately, to experience. How this becomes conscious experience remains as mysterious as ever (which may mean: not at all) but my proposal indicates the direction: It requires B), the awareness of input from different senses, and we require the perception of spatial arrangements as spatial arrangements - rather than encodings of spatial arrangements.

Of course, I was talking about vision only, and the content I proposed applies only to the visual module - but it seems apt to assume that similar considerations would also apply to other perceptual modules (auditory, tactile, olfactory, taste, proprioception, …). Given that in actual human ontogeny our various senses interact, it is a possibility that what I proposed as necessary for the visual module might be supplied by the other sensual input modules - this is not likely, in view of the considerations above, but a possibility to be kept in mind for further investigation.

I conclude that there must be informationally encapsulated content in the visual perceptual module as Fodor had proposed 20 years ago. What remains to be seen is what exactly that content is - and here I suspect Fodor will prove too generous.

# REFERENCES

Chomsky, Noam (1980) "Rules and Representations", *Behavioral and Brain Sciences* 3, 1-15.

Cosmides, L./Tooby, J. (1994) "Origins of Domain Specificity: The Evolution of Functional Organization". In: *Mapping the Mind*, ed. L. Hirschfeld and S. Gelman. Cambridge. Cambridge University Press.

Dretske, Fred (1993) "Conscious Experience". *Mind* 102 (406), 263-283

— (1997) *Naturalizing the Mind.* Cambridge, Mass.: The MIT Press.

Fodor, Jerry A. (1983) *The Modularity of Mind*. Cambridge, Mass.: The MIT Press.

— (2000) *The Mind Doesn't Work that Way: The Scope and Limits of Computational Psychology*. Cambridge, Mass.: The MIT Press.

Gunther, York H. (2003) (ed.) *Essays on Nonconceptual Content*. Cambridge, Mass.: The MIT Press.

Harnad, Stephen (1990). "The Symbol Grounding Problem". *Physica D*, 42, 335-346.

– (1993) "Symbol Grounding is an Empirical Problem: Neural Nets are Just a Candidate Component." In: *Proceedings of the Fifteenth Annual Meeting of the Cognitive Science Society.* NJ: Erlbaum.

Hofstadter, Douglas R. (1979): *Gödel, Escher, Bach: An Eternal Golden Braid*. New York: Basic Books.

Kant, Immanuel (1781) *Kritik der reinen Vernunft* in *Werke,* ed. W. Weischedel, vol. II. Darmstadt: Wissenschaftliche Buchgesellschaft 1956.

Leibniz, Gottfried Wilhelm (1705) *Nouveaux essais sur l'entendement humain*, ed. J. Brunschwig. Paris: Garnier-Flammarion 1966.

Locke, John (1689) *An Essay Concerning Human Understanding*, ed. P. Nidditch. Oxford : Oxford University Press 1975.

Marr, David (1982) *Vision*. New York: Freeman.

McDowell, John (1994) *Mind and World.* 2nd edition with a new introduction. Cambridge, Mass.: Harvard University Press 1996.

Putnam, Hilary (1967) "The 'Innateness Hypothesis' and Explanatory Models in Linguistics", in *Mind, Language and Reality, Philosophical Papers II.* Cambridge: Cambridge University Press 1975, 107-116.

— (1984) "Models and Modules: Fodor's *The Modularity of Mind*" in *Words and Life.* Cambridge, Mass.: Harvard University Press 1994, 403-415.

Pylyshyn, Zenon (1999) "Is Vision continuous with Cognition? The Case for Cognitive Impenetrability of Visual Perception". *Behavioral and Brain Sciences* 22, 341-423.

Raftopoulos, Athanasios (2001) "Is Perception Informationally Encapsulated? The Issue of the Theory-Ladeness of Perception". *Cognitive Science* 25, 423-451.

Raftopoulos, Athanasios/Müller, Vincent C. (2003a) "Deictic Codes, Object-Files and Demonstrative Reference" submitted to *Mind.*

— (2003b) "The Nonconceptual Content of Experience" submitted to *Mind and Language.*

Searle, John (1980) "Minds, Brains and Programs", *Behavioral and Brain Sciences* 3: 417-457.

— (1999) "I Married a Computer" (review of Ray Kurzweil *The Age of Spiritual Machines*). *The New York Review of Books* 08.04.1999, 35-38.

Tye, Michael (2002) "Visual Qualia and Visual Content Revisited". In David J. Chalmers (ed.) *Philosophy of Mind*. New York: Oxford University Press, 447–457.

Wolff, Christian (1713) *Deutsche Logik*, in *Gesammelte Werke*, ed. J. École/J.E. Hofmann/M. Thomann/H.W. Arndt. Hildesheim: Olms 1968-1991, vol. I/1.